\documentclass[letterpaper,aps,showpacs,twocolumn,eqsecnum,floatfix,amsmath,amssymb]{revtex4}

\usepackage{amssymb}
\usepackage{amsmath}
\usepackage{epsfig}
\usepackage{graphicx}
\usepackage{latexsym}
\usepackage{bm,latexsym}
\usepackage{mathrsfs}

\begin{document}


\title{Dynamic transition to spontaneous scalarization in boson stars}

\author{Miguel Alcubierre$^1$}
\email{malcubi@nucleares.unam.mx}

\author{Juan C. Degollado$^1$}
\email{jcdegollado@nucleares.unam.mx}

\author{Dar\'\i o N\'u\~nez$^{1}$}
\email{nunez@nucleares.unam.mx}

\author{Milton Ruiz$^2$}
\email{milton.ruiz@uni-jena.de}

\author{Marcelo Salgado$^1$}
\email{marcelo@nucleares.unam.mx}

\affiliation{$^1$Instituto de Ciencias Nucleares, Universidad Nacional
Aut\'onoma de M\'exico, A.P. 70-543, M\'exico D.F. 04510, M\'exico \\
$^2$ Theoretical Physics Institute, University of Jena, 07743 Jena, Germany}


\date{\today}


\begin{abstract}
We show that the phenomenon of spontaneous scalarization predicted in
neutron stars within the framework of scalar-tensor tensor theories of
gravity, also takes place in boson stars without including a
self-interaction term for the boson field (other than the mass term),
contrary to what was claimed before. The analysis is performed in the
physical (Jordan) frame and is based on a 3+1 decomposition of
spacetime assuming spherical symmetry.
\end{abstract}




\maketitle


\section{Introduction}
\label{sec:introduction}

Scalar-tensor theories of gravity (STT) are alternative metric
theories of gravitation where a spin-0 gravitational degree of freedom
$\phi$ accompanies the usual tensor spin-2 modes (see
Ref.~\cite{Damour92} for a review). In the so-called Jordan frame, the
scalar field $\phi$ couples non-minimally to the curvature while in
the Einstein frame, it couples non-minimally to the ordinary matter
fields. Scalar tensor theories are perhaps the most analyzed and
competitive theories of gravitation after general relativity, the most
prominent example being the well-known Brans-Dicke
theory~\cite{Brans61}. Nevertheless, is only recently that many issues
and theoretical discoveries concerning STT have been settled. In the
cosmological context, STT have been proposed as alternatives to dark
energy in order to explain the accelerated expansion of the Universe
while avoiding the so called {\em coincidence problem} which is
associated with a cosmological
constant~\cite{Perrotta1999,Boisseau2000,Amendola2001,Riazuelo2002,Salgado02,Schimd2005}.
In the astrophysical context, Damour and
Esposito-Far\`ese~\cite{Damour93,Damour96} discovered that neutron
star models (polytropes) within STT may undergo a {\em phase
  transition} that consists in the appearance of a non-trivial
configuration of the scalar field $\phi$ in the absence of sources and
with vanishing asymptotic value. Such configurations are endowed with
a new global quantity termed {\em scalar charge}.  Due to the
similarities of this scalarization process with the spontaneous
magnetization in ferromagnets at low temperatures, these authors
coined the term {\em spontaneous scalarization} (SC) to describe this
phenomenon.  Further analysis~\cite{Damour98,Novak98b,Salgado98}
confirmed that this phenomenon also takes place in realistic neutron
star models and occurs independently of the details of the equation of
state (EOS) used to describe the nuclear matter.

The stability analysis for the transition to SC was first performed by
Harada~\cite{Harada97,Harada98}. It is now understood that SC arises
under certain conditions where the appearance of the scalar field
gives rise to a configuration that minimizes the star's energy (the
ADM mass) with fixed baryon number. This interpretation can in fact
accommodate the Newtonian expectations despite the fact that the
effective gravitational constant decreases during the
transition~\cite{Salgado98}. Following the analogy with
ferromagnetism, the ADM mass plays the same role as the (free) energy
of the ferromagnet, while the baryon mass is the analogue of the
inverse of the temperature.  The order parameter is the scalar charge
which mirrors the magnetization.  One important aspect of this
phenomenon is that it occurs even when the parameters of the theory
satisfy the stringent bounds put by the solar system experiments. The
important point is that SC appears precisely when the asymptotic
(cosmological) value $\phi_0$ of the scalar field vanishes. Therefore,
the phenomenon arises even when the associated effective Brans-Dicke
parameter of the theory is arbitrary large (see Sec.~\ref{sec:discussion}).  
On the other hand, the binary pulsar does put limits on the magnitude of SC. 
In some classes of STT, these bounds restrict the non-minimal coupling to the
curvature. However, the bounds are no so stringent since the couplings
can still be of order unity~\cite{Damour96}.

More recently, studies of neutron star oscillations within STT reveal
that, in addition to the emission of scalar gravitational waves, SC
can also disturb the quadrupolar gravitational radiation as compared
to the corresponding signals in general relativity (GR).  Therefore,
even if the detection channels of scalar gravitational waves are
``switched off'', the detection of gravitational waves of spin-2
coming from these sources might validate STT or put even more
stringent bounds on their parameters. Of course, the direct detection
(or lack) of scalar gravitational waves would also help to
discriminate between several alternative theories. In this regard it
is important to emphasize another striking feature of STT. While GR
predicts only quadrupolar gravitational radiation in the ``far zone'',
STT predicts monopolar gravitational waves~\cite{Maggiore2000}, so
that even in spherical symmetry scalar waves can be emitted.  This is
in part because Birkhoff's theorem does not apply in this case. The
new polarization scalar mode is of {\em breathing} type since it
affects all the directions isotropically~\cite{Maggiore2000}. In
particular, during the SC process in spherical neutron stars this kind
of radiation might be emitted. In fact, using a fully relativistic
spherically symmetric code, Novak~\cite{Novak98b} not only confirmed
the dynamical transition to the scalarization state but also the
emission of such scalar waves.

In a more recent analysis, Whinnett~\cite{Whinnett00} corroborated
that SC can also occur in boson stars (see Ref.~\cite{Jetzer92} for a
thorough introduction to the subject), but only if these are endowed
with a self-interaction. In that work the spacetime was assumed to be
static and spherically symmetric, although the boson stars were only
stationary.

In this paper we want to report that by performing a dynamical
transition to SC in boson stars where the real scalar field $\phi$ is
excited to a non-trivial configuration, we find that the
self-interaction term for the boson field (other than the mass term)
is not required, contrary to what was found by Whinnett. This result
is not the only novelty.  In order to perform our analysis we have
developed several new tools.  Unlike Novak's work, we always work in
the Jordan frame where the interpretation of results has a more direct
physical meaning. To achieve this goal we have constructed a new code
based on a 3+1 approach for STT developed in~\cite{Salgado06} and
adapted to a BSSN formulation~\cite{Shibata95,Baumgarte:1998te}, in
which we have implemented constraint preserving boundary conditions.
In this regard, it is important to mention that STT have been proved
to possess a well-posed Cauchy problem in the Jordan
frame~\cite{Salgado06,Salgado08}. In particular, we shall show
elsewhere~\cite{Alcubierre2010}, that the spherically symmetric
equations used in this paper are strongly hyperbolic and therefore the
Cauchy problem is well-posed as well for this particular case.

The paper is organized as follows: In Section~\ref{sec:STT} we
introduce the Scalar-Tensor Theories and discuss briefly some
properties associated with the Jordan frame.  In
Section~\ref{sec:bosons} we describe the boson-star
model. Section~\ref{sec:numres} describes the numerical results which
include some comments about the scalar-waves emitted, the details of
which will appear elsewhere~\cite{Alcubierre2010}.  Finally,
Section~\ref{sec:discussion} contains the conclusions.


\section{Scalar-tensor theories of gravity}
\label{sec:STT}

The general action for STT with a single scalar field is given by
\begin{eqnarray}
S[g_{ab}, \phi, {\mbox{\boldmath{$\psi$}}}] &=&
\!\! \int \!\! \left\{ \frac{F(\phi)}{16\pi} R
- \frac{1}{2}(\nabla \phi)^2 - V(\phi) \right\} \sqrt{-g} \: d^4 x
\nonumber \\
&+& S_{\rm matt}[g_{ab}, {\mbox{\boldmath{$\psi$}}}] \; ,
\label{jordan}
\end{eqnarray}
with $\phi$ the non-minimally coupled scalar field, and where
${\mbox{\boldmath{$\psi$}}}$ represents collectively the ordinary
matter fields ({\em i.e.} fields other than $\phi$; we use units such
that $c=1=G_0$). For the problem at hand ${\mbox{\boldmath{$\psi$}}}$
will represent the complex scalar field that we use to model the boson
star (see Sec.~\ref{sec:bosons}).

The representation of the STT given by Eq.~(\ref{jordan}) is called
the {\em Jordan frame}\/ representation. The field equations obtained
from the action~(\ref{jordan}) are given by
\begin{eqnarray}
\label{Einst}
G_{ab} &=& 8\pi T_{ab}\,\,\,\,, \\
\label{KGo}
\Box \phi &+& \frac{1}{2}f^\prime R = V^\prime \,\,\,,
\end{eqnarray}
where a prime indicates $\partial_\phi$,
$\Box:=g^{ab}\nabla_a\nabla_b$ is the covariant d'Alambertian
operator, $G_{ab}=R_{ab}-\frac{1}{2}g_{ab}R$, and
\begin{eqnarray}  
\label{effTmunu}
T_{ab} &:=& G_{{\rm eff}}\left(\rule{0mm}{0.5cm} T_{ab}^f
+ T_{ab}^{\phi} + T_{ab}^{{\rm matt}}\right) \; , \\
\label{TabF}
T_{ab}^f &:=& \nabla_a \left( f^\prime \nabla_b \phi \right)
- g_{ab} \nabla_c \left( f^\prime \nabla^c \phi\right) \; , \\
T_{ab}^{\phi} &:=& (\nabla_a \phi)(\nabla_b \phi) - g_{ab}
\left[ \frac{1}{2}(\nabla \phi)^2 + V(\phi)\right ] \; , \qquad \\
\label{Geff}
G_{{\rm eff}} &:=& \frac{1}{8\pi f} \; ,\qquad f:=\frac{F}{8\pi} \; ,
\end{eqnarray}
where $T_{ab}^{\rm matt}$ is the stress-energy tensor for all matter
  fields other that $\phi$.

Using Eq.~(\ref{Einst}), the Ricci scalar can be expressed in terms of
the energy-momentum tensor Eq.~(\ref{effTmunu}). Equation~(\ref{KGo})
then takes the final form
\begin{equation}
\label{KG}
{\Box \phi} = \frac{ f V^\prime - 2f^\prime V - \frac{1}{2} f^\prime
\left( 1 + 3f^{\prime\prime} \right)(\nabla \phi)^2
+ \frac{1}{2} f^\prime T_{\rm matt} }
{f \left(1 + 3 {f^\prime}^2 / 2f \right) } \: ,
\end{equation}
where $T_{{\rm matt}}$ stands for the trace of $T_{ab}^{\rm matt}$.
It is important to stress the fact that, although we have included
here a self-interaction potential $V(\phi)$ for the STT, for the
actual analysis of SC in boson stars that we shall present below it
was not taken into account.

Using a modified harmonic gauge~\cite{Salgado06}, it has been proved
that the previous equations (in vacuum) can be put in a quasi-linear
diagonal hyperbolic form.  Moreover, such equations can also be put in
a full first order 3+1 form, from which one can prove directly their
strong hyperbolic character~\cite{Salgado08}.  Therefore, the Cauchy
problem is well posed in the Jordan frame.

The Bianchi identities directly imply
\begin{equation}
\nabla_{c } T^{c a } = 0 \; .
\end{equation}
However, the use of the field equations leads to the conservation of
the energy-momentum tensor of the matter alone
\begin{equation}
\nabla _{c }T_{{\rm matt}}^{c a } = 0 \; ,
\end{equation}
which implies the fulfillment of the (weak) equivalence principle
({\em i.e.} test particles follow geodesics of the metric $g_{ab}$.)


\section{Boson Stars}
\label{sec:bosons}

As we mentioned in the introduction, the phenomenon of SC was first
discovered in static neutron star models.  Afterwards, the dynamic
transition from the unscalarized to the scalarized state was also
analyzed~\cite{Novak98b}. In this paper, we propose to study a similar
transition but using boson stars instead of neutron stars. In some way
this matter model is even more fundamental than neutron stars in that
one does not have to assume a perfect fluid, but rather a field
theory. It is also simpler since one does not have to deal with the
EOS for the nuclear matter (where uncertainties in the model are
always an issue), as well as the technical difficulties associated
with the numerical simulation of shock fronts.

In a previous investigation, Whinnett~\cite{Whinnett00} constructed
stationary configurations of boson stars within a STT, and showed that
SC was only possible if one included a self-interaction potential for
the boson field.  He carried out the analysis for three different
classes of STT, one of which in fact is almost identical to the one we
use in this paper. In our case, instead of constructing stationary
configurations of scalarized boson stars, we analyse the dynamical
transition from the unscalarized state to the one with
scalarization. The final state of this process correspond in principle
to one of the stationary states that one would found using the method
by Whinnett.  Nevertheless, unlike Whinnett's results, we find SC
without the need of a self-interaction.  The possible explanation for
this will be elucidated in Sec.~\ref{sec:discussion}.  We point out
that boson-star models have been constructed in the past within
STT~\cite{Torres97,Comer97,Torres98a,Torres98b}, but apart from
Whinnett's work none other study has looked at the spontaneous
scalarization of such objects.

We shall consider boson stars described by the following stress-energy
tensor
\begin{eqnarray}
T_{ab}^{\rm matt} &=& \frac{1}{2} \left[\rule{0mm}{0.4cm} (\nabla_a \psi^*)(\nabla_b \psi)
+ (\nabla_b \psi^*)(\nabla_a \psi) \right] \nonumber \\
&-& g_{ab} \left[ \frac{1}{2} |\nabla \psi|^2 + V_\psi(|\psi|^2) \right] \; ,
\end{eqnarray}
where $\psi$ is a complex-valued scalar field that represents the
bosons, $|\nabla \psi|^2:= g^{ab}(\nabla_a \psi) (\nabla_b \psi^*)$
and \mbox{$|\psi|^2:= \psi^*\psi$}.  This model arises from the
Lagrangian ${\cal L}_{\rm matt} = - \sqrt{-g} \left[
  \frac{1}{2}|\nabla \psi|^2 + V_\psi(|\psi|^2) \right ]$.  The
potential $V_\psi(\psi^*\psi)$ will be taken to be of the form
\begin{eqnarray}
V_\psi(\psi^*\psi) = \frac{1}{2} m_\psi^2 \psi^* \psi  \; ,
\label{bospot}
\end{eqnarray}
which includes a mass term but no self-interaction potential (which is
typically associated with a term $\frac{1}{4}\lambda(\psi^* \psi)^2$)
\footnote{At this point a comment on the units is in order. The only
  scale appearing in the STT-boson field theory is
  $m_\psi$. Therefore, the coordinate $r$, the total ADM-boson mass
  and the scalar charge will be all in units of $m_\psi^{-1}$.  When
  $G_0$ and $c$ are restored, the total boson mass is usually given in
  units of $M_p^2/m_b$ where $M_P:=\sqrt{\hbar c/G_0}$ is the Planck
  mass and $m_b:= \hbar m_\psi/c$ is the mass of single bosons; the
  energy density is measured in units $\rho_c:= m_\psi^2 c^4/G_0=
  \rho_P (m_b/M_P)^2$ where $\rho_P:= c^7/(\hbar G_0^2)$ is the Planck
  energy density. Note that $m_\psi^{-1}= l_p (m_b/M_p)$ where
  $l_p=\sqrt{\hbar G_0/c^3}$ is the Planck length.}.

The boson field obeys the Klein-Gordon equation:
\begin{equation}
\label{KGbos}
{\Box \psi} - m_\psi^2 \psi = 0 \; .
\end{equation}

Notice that the problem we wish to study involves two {\em different}\/
scalar-fields, a real-valued field $\phi$ coupled non-minimally to the
curvature, and a complex-valued field $\psi$ (the boson field) coupled
only minimally.

The matter Lagrangian is invariant with respect to a global phase
transformation $\psi \rightarrow e^{\imath q} \psi$ (with $q$ a real
constant). Noether's theorem then implies the local conservation of
the boson number \mbox{$\nabla_a {\cal J}^a=0$}, where the
number-density current is given by \mbox{${\cal J}^a =
  \frac{\imath}{2}g^{ab}\left[\psi\nabla_b\psi^* -\psi^*\nabla_b \psi
    \right]$}. This means that the total boson number
\begin{equation}
 {\cal N} = - \int_{\Sigma_t} \! n_a {\cal J}^a \sqrt{h} \; d^3 x \; ,
\end{equation}
is conserved, where $n^a$ is the normal vector to the spatial Cauchy
hypersurfaces $\Sigma_t$ and $h$ is the determinant of the induced
metric on $\Sigma_t$. The total boson mass is then given by $M_{\rm
  bos}= m_b {\cal N}$, with $m_b:= \hbar m_\psi/c$ the mass of a
single boson (where we have restored the speed of light $c$ and the
Planck's constant $\hbar$).

In fact, when the scalar field $\phi$ acquires a non-trivial value one
can define another global quantity associated to it, which for
asymptotically flat spacetimes is given by
\begin{equation}
 Q_{scal} = \lim_{r \rightarrow \infty} \frac{1}{4\pi}
\int_{S} s_a \nabla^a \phi ds \; ,
\end{equation}
where $s^a$ is the unit outward normal to a topological 2-sphere $S$
embedded in $\Sigma_t$, and $r$ is a radial coordinate that provides
the area of $S$ asymptotically.


\section{Numerical results}
\label{sec:numres}

For the analysis at hand, we shall consider a spherically symmetric
(SS) spacetime with a metric given by
\begin{equation}
\label{SSmetric}
ds^2 = - N^2 dt^2 + A^2 dr^2 + r^2 B^2\left( d\theta^2
+ \sin^2\theta d\varphi^2 \right) \; , 
\end{equation}
where the metric coefficients $(N,A,B)$ are all functions of the
coordinates $t$ and $r$. The scalar-field variables will be functions
of $t$ and $r$ as well. As can be seen from the form
of~\eqref{SSmetric}, we consider a null shift vector. Also, the area
of 2-spheres is given by ${\cal A}= 4\pi r^2 B^2$, which only
coincides with the area coordinates when $B\equiv 1$.

We have constructed a spherically symmetric evolution code based on
the BSSN system of equations, together with a 3+1 formulation of the
STT equations developed in Ref.~\cite{Salgado06}. The details of this
system, including its hyperbolicity properties, will be reported
elsewhere~\cite{Alcubierre2010}.

For the evolution we have also used a generalization of
the Bona-Masso slicing condition~\cite{Bona94b} that led to 
a well behaved hyperbolic system~\cite{Salgado08}.

We have constructed initial data for stationary boson stars like in
GR, by assuming $\psi(t,r)= \Psi(r) e^{\imath \omega \, t}$ and
$\phi(t,r)\equiv 0$, and solving the eigenvalue problem resulting from
Eq.~(\ref{KGbos}) using a shooting method to find $\omega$.  Notice
that $\phi(t,r) \equiv 0$, solves exactly Eq.~\eqref{KG}. The
resulting configuration corresponds to a strictly static spacetime.
In order to study the dynamic transition to SC, we then considered a
small Gaussian perturbation for the scalar field $\phi$. We then
solved the new Hamiltonian constraint for this perturbed initial
data. The momentum constraints are trivially satisfied initially by
assuming a {\em moment of time symmetry} for which the extrinsic
curvature vanishes, as well as the momenta associated with the real
scalar field $\phi$ and the boson field $\psi$.

For the dynamical simulation presented below, we have taken the
non-minimal coupling function to be of the form \mbox{$F(\phi)=8\pi
  f(\phi) = 1 + 8\pi \xi \phi^2$}, with $\xi$ a positive constant. 

\begin{figure}[h!]
\includegraphics[width=0.5\textwidth]{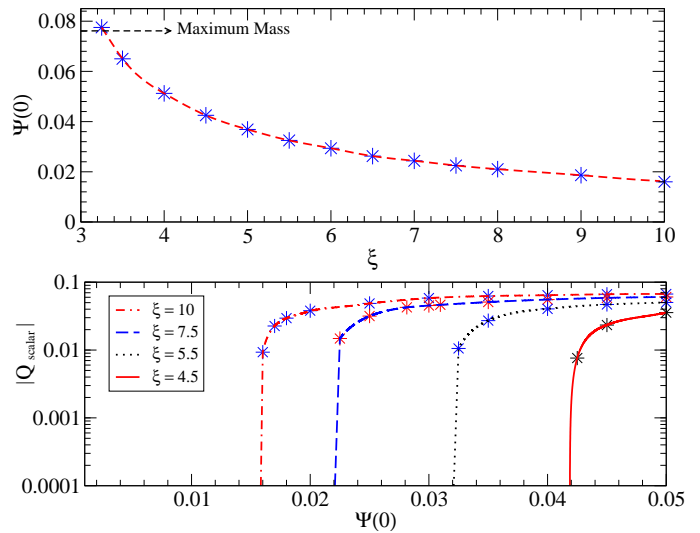}
\caption{(Top) Critical values of central boson field $\Psi_{\rm
    central}$ and non-minimal coupling parameter $\xi$ for spontaneous
  scalarization; (Bottom) Scalar charge at the end of the transition
  to spontaneous scalarization for different values of the
  parameters.}
\label{fig:1}
\end{figure}

Several scenarios can happen depending on the initial
configuration. Figure~\ref{fig:1} (top panel) depicts a curve which
represents the boundary ({\em i.e.} the critical values) of the
transition to SC for different values of the constant $\xi$ and the
central value of the norm of the complex scalar field $\Psi(0)$.
Initial configurations below the critical line are stable with respect
to the perturbations and do not lead to a SC transition.  This means
that for such configurations the scalar field $\phi$ simply radiates
away during the evolution and leaves behind a stable stationary
configuration with a globally null $\phi$.  On the other hand, initial
configurations above the critical line (but below the line marked
``maximum mass'') are unstable, and when perturbed lead to a dynamical
transition to SC where the final configuration is a stationary boson
star endowed with a non-trivial scalar field $\phi$.

The horizontal line marked ``maximum mass'' in Figure~\ref{fig:1}
denotes the maximum mass for stable boson star configurations.  For
boson stars with no perturbation from the scalar field $\phi$ the mass
increases with increasing $\Psi(0)$ until a threshold value
$\Psi(0)_{\rm crit}$ is reached, after which the masses start to
decrease for larger $\Psi(0)$. This value separates two regions, the
values with $\Psi(0)<\Psi(0)_{\rm crit}$ represent stable boson stars,
while the configurations with $\Psi(0)>\Psi(0)_{\rm crit}$ are
unstable and either collapse to a black hole or migrate to a
configuration on the stable branch~\cite{Seidel90}. The maximum mass
is $~0.633\,m_\psi^{-1}$ and corresponds to $\Psi(0)_{\rm crit}=0.076$
or $\sigma(0)_{\rm crit}=0.27$ with the usual normalization
($\sigma=\sqrt{4\,\pi\,}\,\Psi$). Similar behavior is expected to
occur in STT~\cite{Alcubierre2010}.

\begin{figure}[h!]
\epsfxsize=50mm 
\includegraphics[width=0.52\textwidth]{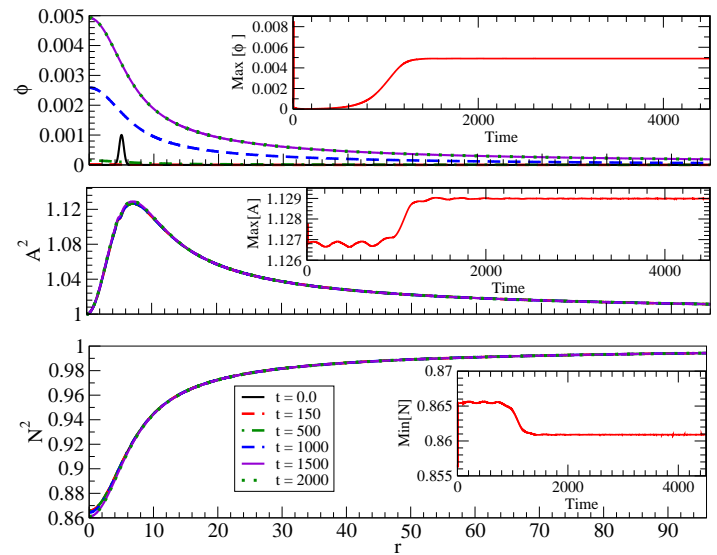}
\caption{NMC scalar field $\phi$ and metric components at different
  times for the simulation with $\xi=6$ and $\Psi(0)=0.03$.}
\label{fig:metric}
\end{figure}

Figure~\ref{fig:metric} depicts a series of snapshots taken at
different times during the evolution for the unstable case with $\xi=
6$ and $\Psi(0)= 0.03$. Note from this figure that initially the
scalar field $\phi$ almost vanishes, whereas at the end of the
evolution it settles down into a stationary configuration which
results in a non-trivial profile that interpolates between a finite
value $\phi(t_{final},r=0)$ at the center of the boson star and a
vanishing value $\phi(t_{final},r_\infty)$ asymptotically. The final
stationary configuration is also characterized by the appearance of a
global {\em scalar charge} (see Figure \ref{fig:1} bottom panel).

\section{Discussion}
\label{sec:discussion}

Boson stars are stable self-gravitating configurations of a complex
scalar field, and as such they could in principle exist in nature. If
the bosons are light ($m_b:=\hbar m_\psi/c \sim {\rm eV}/c^2$) one
expects the mass of the boson star to be of order $M \sim
10^{20}\,{\rm kg}$, which is in the mass range of some asteroids. On
the other hand, for heavy bosons ($m_b\sim 100 {\rm GeV}/c^2$) the
boson star mass turns out to be much lower ($M \sim 10^{9}\,{\rm
  kg}$). However, by including self-interactions it is possible to
increase the mass of the boson star to the order of the solar
mass~\cite{Colpi86}. It has also been speculated that super-massive
boson stars instead of black holes could be at the center of
galaxies~\cite{Torres2000}.  If boson stars actually exist in nature,
they could serve as natural ``laboratories'' to test different
alternative theories of gravity.

In this work we have used a scalar tensor theory of gravity to study
dynamical simulations of boson stars. We have found that, just like in
the case of neutron stars, boson stars can also undergo a spontaneous
scalarization process.  We have analyzed this transition in a
dynamical fashion using a fully relativistic code in spherical
symmetry.  Unlike previous studies of stationary boson star
configurations, we have found that self-interactions are in fact not
required to produce scalarization.  We have implemented a STT which is
very similar to the one studied by Whinnett, where scalarization was
not found without self-interaction. However, he used a value
$\xi_W=1$~\footnote{Whinnett's notation differs from hours in the
  following: his non-minimally coupling (NMC) field $\chi$ is related
  to ours by $\chi=\sqrt{16\pi}\phi$ and his NMC $\xi_W$ is $\xi_W=
  \xi/2$. Moreover his complex boson field $\Psi_W$ relates to ours by
  $\Psi_W= \psi/\sqrt{2}$.} (corresponding to our $\xi=2$) which are
actually in the region of Figure~\ref{fig:1} where scalarization does
not ensue. He also used $\xi_W=2$ (ours $\xi=4$) where our results of
Figure~\ref{fig:1} show that there is a small region where
scalarization is found. Actually from Whinnett's figure 4 is not clear
that scalarization is not found for $\xi_W=2$ since his coupling
parameter is not null. By increasing the value of the non-minimal
coupling parameter $\xi$ one can reach a threshold where the
energetically favorable configurations are those with the presence of
a non-trivial scalar field $\phi$. Note from Figure~\ref{fig:1} that
the larger the value of the parameter $\xi$, the lower the central
energy density required to produce the transition to scalarization.

At this point is perhaps appropriate to mention that the effective
Brans-Dicke parameter given by \mbox{$\omega_{\rm BD}=
  f/(f')^2|_{\phi_0}$} (where $\phi_0$ is the asymptotic
(cosmological) value) takes the explicit form \mbox{$\omega_{\rm BD}=
  (1+ 8\pi \xi \phi_0^2)/[32\pi(\xi\phi_0)^2]$}.  Therefore, as the
spontaneous scalarization ensues with $\phi_0\rightarrow 0$ with a
finite value of $\xi$, one has $\omega_{\rm BD}\rightarrow \infty$
which obviously passes the constraints imposed by the Cassini probe
$|\gamma-1| \lesssim 2.3\times 10^{-5}$~\cite{Bertotti2003}, where
$\gamma$ is the post-Newtonian parameter which in terms of
$\omega_{\rm BD}$ is given by $\gamma=(\omega_{\rm BD} + 1)/ (\omega_{\rm BD} + 2)$,
implying $\omega_{\rm BD} \gtrsim 4.3\times 10^4$. As was already
mentioned in the introduction, the phenomenon of spontaneous
scalarization can therefore appear independently of the bounds imposed
on $\omega_{\rm BD}$ by the solar system experiments.

Though here we have focused mainly on such a phenomenon and presented
only our main results, in~\cite{Alcubierre2010} we will present a more
detailed explanation of our code and methods, together with a
systematic study of the phenomenon.  Also, in a future work we will
analyse the collapse of a scalarized boson star to a black hole, and
the corresponding emission of scalar gravitational waves.


\acknowledgments

This work was supported in part by CONACyT grants
SEP-2004-C01-47209-F, and 149945, by DGAPA-UNAM grants IN119309-3 and
IN115310. MR was also supported by DFG grant SFB/Transregio 7
“Gravitational Wave Astronomy”.


\bibliographystyle{bibtex/apsrev}
\bibliography{bibtex/referencias}


\end{document}